\documentstyle[psfig,referee]{laa}
\thesaurus{12.03.4; 12.12.1; 11.06.1}
\title{Analytical relations concerning the collapse time in hierarchically clustered cosmological models\\}
\author{M. Gambera\inst{1,2},~  A. Pagliaro\inst{1}} 
\offprints{M. Gambera E-mail mga@sunct.ct.astro.it}
\institute{$^1$ Istituto di Astronomia dell'Universit\`a di Catania, Viale A.Doria, 6 - I 95125 Catania, ITALY \\
$^2$ Osservatorio Astrofisico di Catania and CNR-GNA, 
Viale A.Doria, 6 - I 95125 Catania, ITALY}
\date{}
\begin{document}
\maketitle
\begin{abstract}
By means of numerical methods, 
we solve the equations of motion for the collapse 
of a shell of baryonic matter, made of galaxies and substructure 
falling into the central regions of a cluster of galaxies, 
taking into account the effect of the dynamical 
friction. The parameters on which the dynamical friction mainly 
depends are: the peaks' height, the number of peaks inside a protocluster 
multiplied by the correlation function evaluated at the origin, 
the filtering radius and  
the nucleus radius of the protocluster of galaxies.
We show how the collapse 
time $\tau$ of the shell depends on these parameters.
We give a formula that links the dynamical friction 
coefficient $\eta$ to the parameters mentioned above and 
an analytic relation between the collapse time and $\eta$. 
Finally, we obtain an analytical relation between $\tau$ and
the mean overdensity ${\overline \delta}$ within the shell.
All the analytical relations that we find are in excellent agreement with 
the numerical integration.  
\keywords{cosmology: theory-large scale structure of Universe - galaxies: formation}
\end{abstract}

\section{Introduction}

\noindent
The problem of the formation and evolution of clusters of galaxies has 
been one of the crucial topics of the last years (see {\it e.g.}  
Ryden \& Gunn 1987, Colafrancesco et al. 1989, Antonuccio-Delogu 1992, 
Kaiser 1993, Colafrancesco \& Vittorio 1993, Croft \& Efstathiou 1994, 
Dutta \& Spergel 1994, Mosconi et al. 1994, Sutherland \& Dalton 1994, 
Efstathiou 1994 and Colafrancesco et al. 1995). It is well known that the 
formation of cosmic structures is strictly related to the evolution of 
the density perturbations: in the present {\it paradigm} of 
structure formation, it is generally assumed that cosmic structures 
of size $\sim$ R form preferentially around the local maxima of the 
primordial density field, once it is smoothed on the filtering scale 
$R_{f}$. These linear density fluctuations eventually evolve towards the 
nonlinear regime under the action of gravitational instability; they detach
from the Hubble flow at {\it turn around} epoch $ t_{m}$, given by: 
\begin{equation}
t_{m} = \left[\frac{ 3 \pi}{32 G \rho_{b}} ( 1 +\overline{\delta})
\right]^{1/2} (1+z)^{3/2}
\end{equation}
where $ \rho_{b} $ is the mean background density, $z$ is the redshift and
$\overline{\delta}$ is the mean overdensity within the nonlinear region. After the {\it turn around} epoch, the fluctuations  
start to recollapse when their overdensity defined by
\begin{equation}
\delta({\bf x}) \equiv \frac{ \rho ({\bf x}) - \rho_{b}}{ \rho_{b} }
\end{equation}
reaches the value
$\overline{\delta}=1$.\\
The evolution of the density fluctuations is described by the
power spectrum, given by:
\begin{equation}
{\cal P}( k) \equiv \langle |\delta_{{\bf \scriptstyle k}}|^{2} \rangle 
\end{equation}
with:
\begin{equation}
\delta_{{\bf \scriptstyle k}} \equiv
\int d^{3} k \, e^{-i {\bf  \scriptstyle k x}} \, \delta({\bf x})
\end{equation}

Since the density field depends on the power spectrum, which in turn depends 
on the matter that dominates the universe, the mean characteristics of the 
cosmic structures depend on the assumed model. In this context the most 
successful model is the biased Cold Dark Matter  (hereafter {\sc
CDM}) (see {\it e.g.} Kolb $ \&$ Turner 1990; Peebles 1993; 
Liddle $\&$ Lyth 1993) based on a scale invariant spectrum of 
density fluctuations growing under gravitational instability. 
In such scenario the formation of the structures occurs 
through a  "{\it bottom up} " mechanism.
A simple 
model that describes the collapse of adensity perturbation is that 
by Gunn \& Gott (1972, hereafter {\sc GG72}). The main assumptions of
this model are:  {\bf (a)} the symmetry of the 
collapse is spherical;  {\bf (b)}  the matter distribution is uniform in that 
region of  space  where the density exceeds the background
density; {\bf (c)} no tidal interaction exists with the external 
density perturbations and {\bf (d)}  there is no substructure (collapsed 
objects having sizes less than that of main perturbation). 

Point {\bf (d)}  is in 
contradiction to the predictions of {\sc CDM} models. 
It is well known that in a {\sc CDM} Universe, an 
abundant production of substructures during the evolution of the fluctuations
is predicted.

The problem of the substructures in a {\sc CDM} Universe and his consequences 
on structure formation 
have been widely studied in previous papers (see {\it e.g.}
Antonuccio-Delogu 1992, hereafter {\sc A92}; 
Antonuccio-Delogu \& Atrio-Barandela 1992, hereafter {\sc AA92};
Antonuccio-Delogu \& Colafrancesco 1994, hereafter {\sc AC94}, 
Del Popolo et al. 1996, Gambera 1997 and Del Popolo \& Gambera 1997 hereafter {\sc DG97}).
Being a rather recent topic in cosmology much work has still to be done.\\
The presence of substructure is very important for the 
dynamics of collapsing shells of baryonic matter made of galaxies and 
substructure of $ 10^{6} M_{\odot} \, {\scriptstyle \div} \,
10^{9} M_{\odot}$, falling into 
the central regions of a cluster of galaxies. As shown by Chandrasekhar \& 
von Neumann (1942, hereafter {\sc CvN42}; 1943),
in the presence of substructure it is necessary 
to modify the equation of motion:
\begin{equation}
\frac{d^{2} r} {dt^{2}} = - \frac{GM}{r^{2} (t)}  
\end{equation} 
since the graininess 
of mass distribution in the system induces dynamical 
friction that introduces 
a frictional force term. \\
Adopting the notation of {\sc GG72} (see also their Eqs. 6 and 8) and 
remembering that $T_{c0}/2$ is the collapse time 
of a shell of baryonic matter in
the absence of dynamical friction ({\sc GG72}), one can write:
\begin{equation}
T_{c0} = \frac{\pi \bar{\rho_i} \rho_{ci}^{1/2} } {H_i ( \bar{\rho_i} - \rho_{ci} )^{3/2} }
\end{equation}
where $ \rho_{ci} $ is the 
critical density at a time $ t_{i} $ and $ \bar{\rho_i} $ is 
the {\it average density} inside $ r_{i}$ at $ t_{i}$.
The equation of motion of a shell of baryonic matter in presence of 
dynamical friction (Kandrup 1980, hereafter {\sc K80}; Kashlinsky 1986 
and {\sc AC94}), 
using the dimensionless time variable  $ \tau = \frac{t}{T_{c0}} $,  
can be written in the form:
\begin{equation}
\frac{d^{2} a}{d \tau^{2} }= -
\frac{4 \pi G \rho_{ci}( 1+\overline{\delta_{i}})}{a^{2}(t)} T_{co}^{2}-
\eta T_{c0} \frac{ d a }{ d \tau}  
\label{eq:p}
\end{equation}
where $ \overline{ \delta_{i}}$ is the overdensity within $ r_{i}$, 
$ \eta$ is the coefficient of dynamical friction and $ a(r_{i},t)$ is the 
expansion parameter of the shell (see {\sc GG72} Eq. 6), 
that can be written as: 
\begin{equation}
a( r_{i}, t) = \frac{r(r_{i},t)}{ r_{i}} 
\end{equation} 
Supposing that there 
is no correlation among random force and their derivatives, we have: 
\begin{equation}
\eta = \frac{\int d^{3} F W(F) F^{2} T(F) }{ 2 \langle v^{2} \rangle }
\label{eq:q}
\end{equation}    
({\sc K80}), where $ T(F)$ is the average {\it "duration"} of a random 
force impulse of magnitude $ F$, $ W(F)$ is the probability distribution 
of stochastic force (which for a clustered system is given in Eq. 37 of
{\sc AA92}).\\
{\sc DG97} solved Eq. (\ref{eq:p}) numerically,
showing qualitatively how the
expansion parameter $ a(\tau)$ depends on the dynamical friction coefficient 
and how $ \tau $ changes in the presence of dynamical friction,   
but without undertaking a more complete study of the dependence on
the parameters.\\
The plan of this paper is as follows. In Sect. 2 we show how $ \tau$ 
depends on the peaks' height $\nu_{c}$, on the parameter $\Xi$ that we
define there as the correlation function at the origin multiplied by
the total number of peaks inside a protocluster, on the 
filtering radius $R_{f}$ and on the nucleus radius of the protocluster
$r_0$. 
A more detailed description of these parameters is given below. 
In Sect. 3 we give an analytical relation between the dynamical 
friction coefficient and the collapse time:
\begin{equation}
\tau =  f_1(\eta)
\end{equation} 
In Sect. 4 we give a semi-analytical relation that links the dynamical 
friction coefficient with the parameters on which it depends:
\begin{equation}
\eta = f_2 (\nu_c,R_f,r_0,\Xi)
\end{equation}
Linking these two relations we also find the dependence of the 
dimensionless collapse time $ \tau$ on the parameters used:
\begin{equation}
\tau = f_1 \circ f_2 \equiv f_1 [ f_2 (\nu_c,R_f,r_0,\Xi)]
\end{equation}
In Sect. 5 we give a semi-analytical relation between $\tau$ and 
${\overline \delta}$ and $\eta$:
\begin{equation}
\tau = f_3 ({\overline \delta},\eta)
\end{equation}
Finally, in Sect. 6 we summarize our results and comment on their 
possible implications.

\section{The collapse time} 

{\sc DG97} showed how the expansion parameter $ a(\tau)$ depends on the dynamical
friction, solving Eq. (\ref{eq:p})  
by means of a numerical method  
but not taking into account  
the parameters on which $ \eta$ depends.\\
Here we examine how the dynamical friction  coefficient
$ \eta$ varies according to the parameters and how the collapse
time depends on them. We consider Eq. (\ref{eq:q}) and
the functions  $ T(F)$ and $ W(F)$, the former given
by Chandrasekhar (1943), the latter (for 
clustered system) given by the so-called Holtsmark law 
(Holtsmark 1919):\\
\begin{eqnarray}
W(F) &= &{\frac{2F}{\pi}}  \int_0^{\infty} dk k  \sin(kF)   
\exp \left[ - {\frac{3}{2}} N \left( {\frac{Gm_{typ}}{r_{typ}^2}} k \right)^{3/2} 
\right]   \nonumber \\  
& = & 4 \pi \left| {\bf F} \right| W \left( \left| {\bf F} \right| \right) 
\label{eq:r} \\
\nonumber
\end{eqnarray}
\noindent
where $W(F)dF$  is the probability for a test particle of experiencing a force in the
range $F {\scriptstyle \div} F+dF$, $N$ is the total number of particles, 
$m_{typ}$ is a typical particle mass and $r_{typ}$ 
is a typical distance among the particles.\\
\noindent According to {\sc K80}:
\begin{equation}
W( \left| {\bf F} \right| ) 
= {\frac{1}{2 \pi^2 \left| {\bf F} \right| }}
\int_0^{\infty} dk k \sin( k  \left| {\bf F} \right| ) A_f (k) 
\end{equation}
\noindent with:
\begin{equation}
A_{f}(k) = \lim_{N \to \infty}  A_{N} ({\bf k})  
= A_{\it nor}  
\exp ({\cal F})
\label{eq:s}  
\end{equation}
The original expression for ${\cal F}$ given by {\sc K80} has been modified by
{\sc AA92} to take into account clustering, and turns out to be given by: 
\begin{equation}
{\cal F} =
\left[ - \left\{  A_{\it uncl}(k)  +  A_{\it cl} (k) 
\left[ 1 + {\frac {\Sigma_{\it cl}(k)}{2 A^{\prime}_{\it 2,cl}}} \right] \right\} \right] 
\nonumber
\end{equation}
where $A_{\it nor}$, $ A_{\it cl}(k)$, $A_{\it uncl}(k)$,  $A^{\prime}_{\it 2,(cl)}(k)$ 
and $ \Sigma_{\it cl}(k)$ are given  in {\sc AA92} respectively by the
Eqs. (21), (29), (31), (32) and (36).\\
Here, we want to remind that $ \Sigma_{\it cl}(k)$ is a linear function of the 
correlation function $ \xi(r)$ and that the general expression for
$ \Sigma_{\it cl}(k)$, adopting the notation of {\sc AA92} is given by:\\
\begin{eqnarray}
\Sigma_{\it cl}( {\bf k}) & = &
\int_{\Re^3} d^3 {\bf r}_1  
\int_{\Re^3} d^3 {\bf r}_2   
\exp \left(   i  {\frac{G m_{\it av} {\bf k} {\bf \hat{r}}_1}{r_1^2}} \right)  \nonumber \\
& \cdot & 
\exp \left( i {\frac{G m_{\it av} {\bf k} {\bf \hat{r}}_2}{r_2^2}} \right)
\tau_{\it cl} ({\bf r}_1)
\tau_{\it cl} ({\bf r}_2) 
\xi( \left| {\bf r}_1 - {\bf r}_2 \right| ) \label{eq:t} \\
\nonumber
\end{eqnarray}

\noindent
where masses are measured in units of solar mass $ M_{\odot}$ and distances in 
$h^{-1}$ Mpc, so that $ k$ will be measured in unit of $h^{-1}$(Mpc)$^{2}/GM_{\odot}$,  with 
$m_{av}$ the average mass of the substructure. Since we have 
$ m_{av} {\geq} ~ 10^{6} M_{\odot} >> 1 M_{\odot}$   for all the cases  
that we consider in this paper,
we can adopt the asymptotic expansion of 
$ \Sigma_{\it cl}$ demonstrated in the appendix of {\sc AA92}:
\begin{equation}
\Sigma_{\it cl}(k) \propto \xi(0) 
\end{equation}
(see {\sc AA92} for details). $ \Sigma_{\it cl}$ does not  depend on $ \xi(r)$, 
but only on $ \xi(0)$ (that is $ \xi(r)$ calculated at the origin).\\
We solve Eq. (\ref{eq:q})  and the other equations related
for a outskirts shell of baryonic matter with
$ \overline{\delta} = 0.01$ inside the spherical regions (protocluster),
for different values of $ \nu_{\it c}$,  $ R_{\it f}$,  $ r_{\it 0}$
and  $ \Xi$ (where $ \Xi \equiv N_{tot} \cdot \xi(0)$), the latter quantity 
being better defined in Sect. 4. \\
After having determined $ \eta$ solving numerically
Eq. (\ref{eq:q}),  we get $ \tau$ as a function of 
$ \nu_{\it c}$,  $ R_{\it f}$,  $ r_{\it 0}$ and $ \Xi$
solving Eq. (\ref{eq:p}). We perform these calculations 
for different set of values of $ \nu_{\it c}$,  $ R_{\it f}$,  
$ r_{\it 0}$ and  $ \Xi$ inside the 
following intervals:
\begin{center}
\begin{eqnarray}
1.2 \leq \nu_{c} \leq 3.2 \nonumber \\
0.5 ~h^{-1}{\rm Mpc} \leq ~r_{0} ~\leq 10 ~h^{-1}{\rm Mpc} \nonumber \\
10^{-3} ~h^{-1}{\rm Mpc} ~\leq R_{f} ~\leq 1 ~h^{-1}{\rm Mpc} \\ 
20 \leq  \Xi \leq 2 \cdot 10^{3} \nonumber \\
\nonumber\label{eq:dueaggiunta}
\end{eqnarray}
\end{center}
The results that we have obtained are shown in 
Figs. $1 {\scriptstyle \div} 4$.
\begin{figure}[ht]
\psfig{file=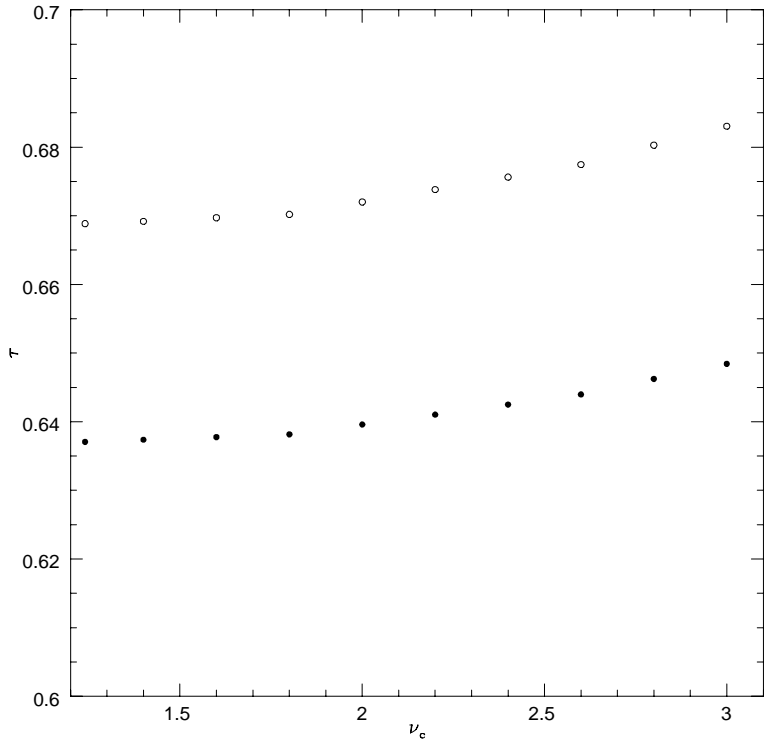,width=11cm}
\caption[h]{ Collapse time $ \tau$
of a shell of matter made of galaxies and substructure 
when dynamical friction is taken into account, 
versus $ \nu_{\it c}$
We assume a nucleus radius of $r_{0} = 1  h^{-1} $Mpc  
and a filtering radius $ R_{f} = 0.74 h^{-1} $Mpc.
{\it Open circles:} $ \Xi = 10^{3}$; {\it filled circles:} $ \Xi 
= 10^{2}$.}
\end{figure}
Before commenting upon the figures, we want to remark that the dependence 
of $ \tau$ on $ \overline{\delta}$ is  qualitatively shown in Fig. 5 
by {\sc AC94}. 
We observe that for $ \overline{\delta} > 10^{-2}$ 
the collapse time in the presence of a dynamical 
friction is always larger than in the unperturbed case but the 
magnitude of the deviation is negligible for larger $ \overline{\delta}$, 
whilst for $ \overline{\delta}  \leq 10^{-2}$ the deviations 
increase steeply with lower $ \overline{\delta}$. Then, having considered
$ \overline{\delta} = 0.01$, the estimation we get for $ \tau$  in  
Sect. 4 must be considered as a lower limit.\\
In Fig. 1 we show the collapse time in the presence of dynamical
friction, versus the peaks'  height, for different values of $ \Xi$. 
In this picture, we  show how $ \tau$ grows for larger values of $ \nu_{\it c}$
and for larger values of $ \Xi$. Similarly,  in Fig. 2  we note 
how $ \tau$ increases for larger values of $ \nu_{\it c}$ and of $ R_{f}$.
\begin{figure}[ht]
\psfig{file=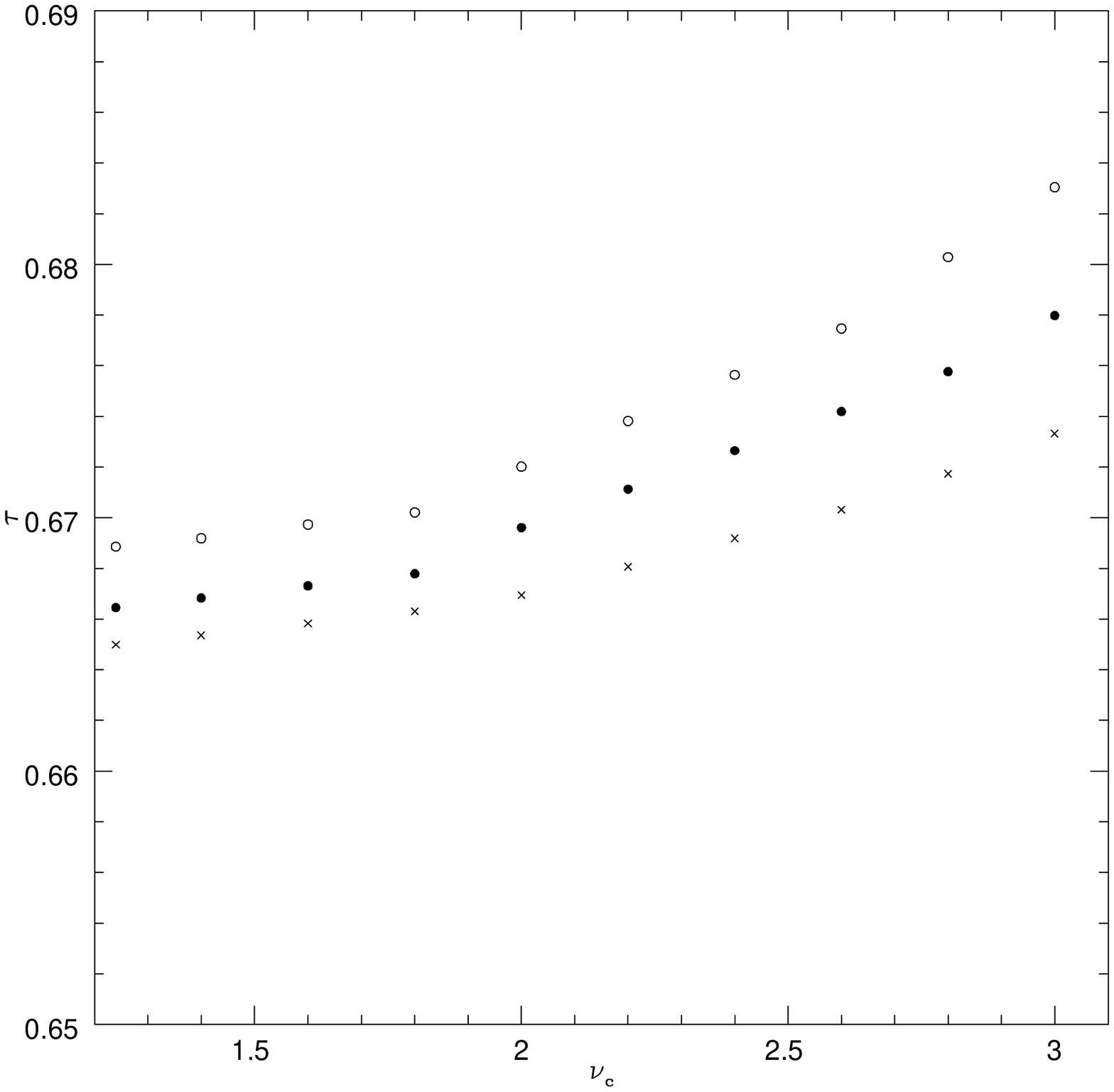,width=11cm}
\caption[h]{Collapse time $ \tau$
of a shell of matter made of galaxies and substructure 
when dynamical friction is taken into account,
versus $ \nu_{\it c}$
We assume a nucleus radius of $r_{0} = 1  h^{-1} $Mpc 
and a fixed correlation $ \Xi = 10^{3}$. 
{ \it Open circles:} $ R_{f} = 0.74 h^{-1}$ Mpc; {\it filled circles:}
$ R_{f} = 0.65 h^{-1}$ Mpc; {\it crosses:} $ R_{f} = 0.55 h^{-1}$ Mpc.}
\end{figure}
The slope of the curves 
confirm our prevision on the 
behaviour of the collapse of a shell of baryonic matter falling into the 
central regions of a cluster of galaxies in the presence of dynamical friction:
the dynamical friction slows down the collapse (as {\sc DG97} had already 
shown) and the effect, as we are showing in Figs. 1 and 
2, increases as $\Xi$, $ R_{\it f}$, $\nu_{\it c}$ grow.\\
Here we want to remind that we are considering only the peaks of the local
density field with central height $ \nu$ larger than a critical 
threshold $ \nu_{\it c}$. This latter quantity is chosen to be the 
threshold at which $ r_{\it peak}$ ($ \nu \geq \nu_{\it c}$) $<<$ $ l_{\it 
av}$ where $ r_{\it peak}$ is the typical size of the peaks and $ l_{\it 
av}$ is the average peak separation (see also Bardeen et al. 1986).\\
\begin{figure}[ht]
\psfig{file=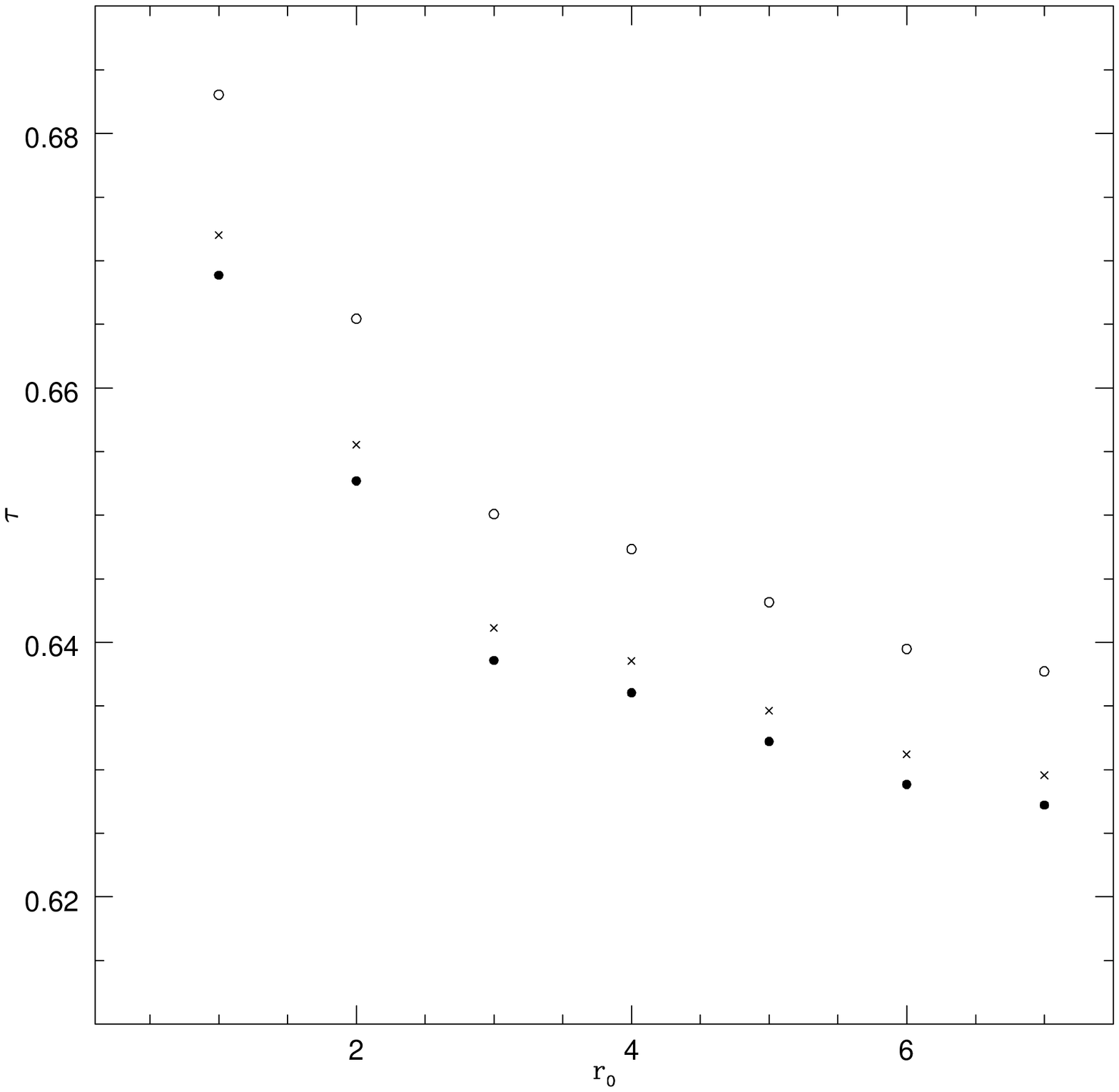,width=11cm}
\caption[h]{Collapse time $ \tau$
versus $ r_{0}$. 
We assume 
a filtering radius $ R_{f} = 0.74 h^{-1}$ Mpc.
and a total number of peaks of substructure $ \Xi = 10^{3}$. 
{ \it Open circles:} $ \nu_{c} = 3 $; {\it crosses:} $ \nu_{c} = 2$; 
{\it filled circles:} $ \nu_{c} = 1.24$.}
\end{figure}
\begin{figure}[ht]
\psfig{file=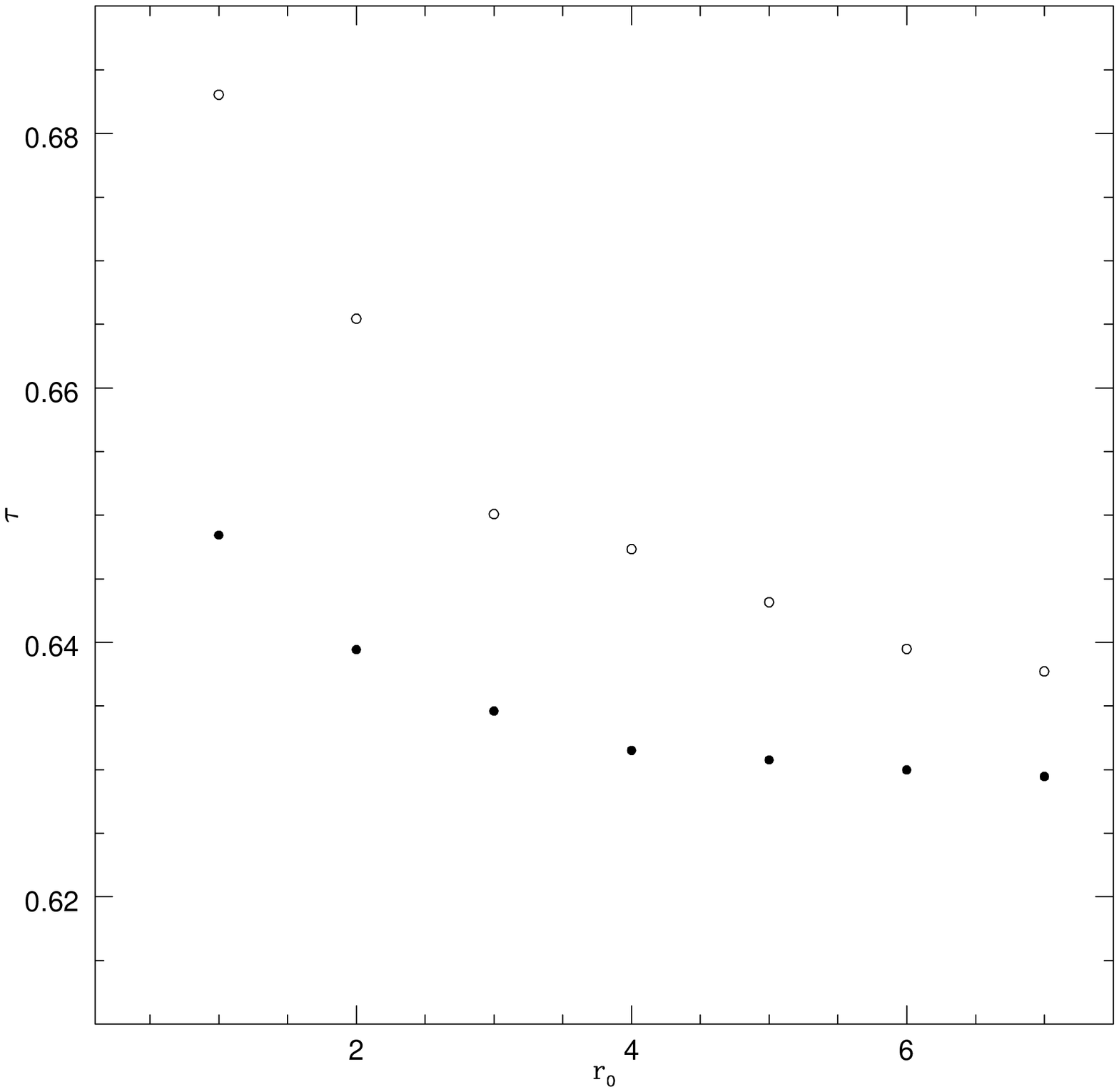,width=11cm}
\caption[h]{ Collapse time $ \tau$
of a shell of matter made of galaxies and substructure 
when dynamical friction is taken into account,
versus $ r_{0}$
We assume 
a filtering radius $ R_{f} = 0.74 h^{-1}$ Mpc 
and a peaks'  height $ \nu_{\it c} = 3$.
{\it Open circles:} $ \Xi = 10^{3}$; {\it filled circles:} $ \Xi 
= 10^{2}$.}
\end{figure}
In Figs. 3 and 4 we show how the collapse time varies with the
nucleus radius of the protocluster $ r_{0}$.
Note how  $ \tau$ grows  as $ r_{0}$ decreases: the smaller 
the nucleus of the protocluster, the larger the time of collapse 
in the presence of dynamical friction; besides we show how this 
effect increases for larger values  of both  $ \nu_{\it c}$ and $ \Xi$.

\section{How the collapse time depends on the dynamical friction} 

Like {\sc DG97}, we have solved  Eq. (\ref{eq:p}) numerically 
and calculated the collapse time $ \tau$ for different values of 
the dynamical friction coefficent $ \eta$. 
However, {\sc DG97} 
showed  the dependence of $ \tau$ on $ \eta$  qualitatively, whilst  
we have 
found also an analytic relation $ \tau = f_{1}( \eta)$ that links
$ \tau$ with $ \eta$ in the range of values $ 0 \leq  \eta \leq 3.1$ 
(see also Gambera 1997).
The results of our calculations are shown in Fig. 5, where we report
the collapse time $ \tau$ versus the dynamical friction coefficient.
\begin{figure}[ht]
\psfig{file=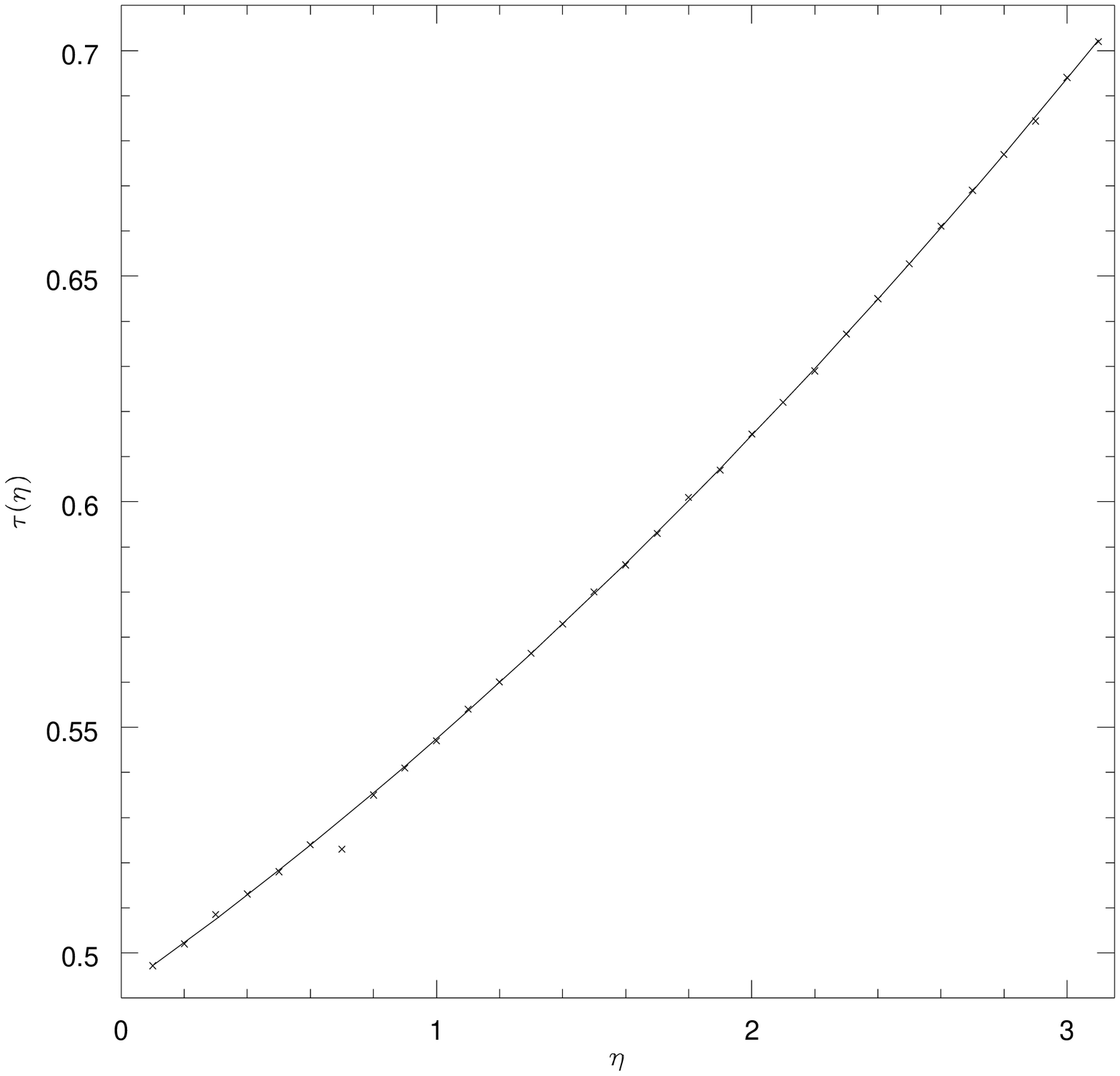,width=11cm}
\caption[h]{Collapse time $ \tau$
of a shell of matter made of galaxies and substructure 
when dynamical friction is taken into account,
versus $ \eta$.}
\end{figure}
We have determined the function $\tau=f_1(\eta)$ through a least squares 
fit method.
Our best solution is given by:
\begin{equation}
\tau = a_1 \eta^2 + a_2 \eta +a_3  
\label{eq:c}
\end{equation}
with: 
\[
\begin{array}{l}
a_1 = 0.00595923, \\
a_2=0.0493460, \\
a_3 = 0.492164, \\
\end{array}
\]
\noindent
Eq. (\ref{eq:c}) and our results for
$0 <  \eta < 3.1$ from the numerical integration of  Eq. (\ref{eq:p})
are in  excellent agreement. As a matter of fact,
the correlation coefficient value obtained 
is: $r^2=1.000$.\\

\section{How the collapse time depends on the parameters} 

As already shown by {\sc AC94}, it is possible to write down the dynamical friction coefficient $\eta$ as the sum of two terms:
\begin{equation}
\eta =\eta_0 + \eta_{cl}  
\label{eq:asa}
\end{equation}
where $\eta_0$ is the coefficient of dynamical friction of an unclustered
distribution of field particles whilst $ \eta_{cl}$ takes into account the effect of clustering. We rewrite Eq. (\ref{eq:asa})  as:
\begin{equation}
\eta =\eta_{0} + \eta_{cl} = \eta_{0} \left( 1 + 
\frac{\eta_{cl}}{\eta_{0}}\right) 
\label{eq:asb}
\end{equation}
\noindent 
where, as demonstrated by {\sc A92}, the ratio  $ \frac{\eta_{cl}}{\eta_0}$ 
depends only on $ \Xi$ and $ r_{0}$ whilst $ \eta_{0}$ is given by {\sc A92}: 
\begin{equation}
\eta_{0}  =  {\frac {3.33}{\pi}}  
\frac{\left[G \langle m \rangle_{av}  p( \gamma, \nu_{c}) c_{p}\right]^{1/2}}
{{\langle n \rangle_{av} R_{cl}^3}}  \log ( 1.36 \langle n \rangle_{av}^{2/3} 
R_{cl}^2 \pi^{2/3})\frac{1}{a^{3/2}}
\label{eq:asc} 
\end{equation}
\noindent 
and, for a fixed value of $\overline{\delta}$, depends only on $\nu_c$ and $R_f$.
In the previous formula, $c_p$ is defined in Bower (1991), 
whilst $ p( \gamma, \nu_{c}),  \langle m \rangle_{av} $ and $ 
\langle n \rangle_{av} $ are defined in {\sc AC94}.
\noindent
Therefore, we rewrite Eq. (\ref{eq:asb})  as:
\begin{equation}
\eta  =  \eta_0(\nu_c,R_f) \left( 1 + \frac{\eta_{cl}}{\eta_0}(r_0,\Xi ) \right) \end{equation}
\noindent 
where the parameters on which $\eta$ depends are the following:
\begin{itemize}
\item $ \; \nu_c$ is the peaks' height;
\item $ \; R_f$ is the filtering radius;
\item $\; r_0$  is the parameter of the power-law density profile. A 
theoretical work (Ryden 1988) suggests that the density profile inside a protogalactic dark matter
halo, before relaxation and baryonic infall, can be approximated by a power-law:
\begin{equation}
\rho(r) = \frac{ \rho_0 r_0^p} {r^p}
\end{equation}
where $p \approx 1.6$ on a protogalactic scale. 
\item $\; \Xi \,$ is the product $N_{tot} \cdot \xi (0)$ where $N_{tot}$ is the
total number of peaks inside a protocluster and $\xi(0)$ is the correlation function calculated in $r=0$.
{\sc AA92} have demonstrated that in the hypothesis $m_{av} \gg 1 M_{\odot} $, 
where $m_{av}$ is the average mass of the subpeaks, the dependence of 
the dynamical friction coefficient on $N_{tot}$ and $\xi(r)$ may be  
expressed as a dependence on a single parameter that we define as:
\begin{equation}
\Xi \equiv N_{tot} \cdot  \xi(0)
\end{equation}
\end{itemize}
\noindent
The analytic relation $\eta=f_2 (\nu_c,R_f,r_0,\Xi)$, that links the 
dynamical friction coefficient with the parameters on which it depends, 
can be rewritten as the product of two functions: 
\begin{equation}
\eta=f_2 (\nu_c,R_f,r_0,\Xi) = f'_2 (\nu_c,R_f) \cdot f''_2 (r_0,\Xi)
\label{eq:asd}
\end{equation}
where
\begin{equation}
\eta_0 = f'_2 (\nu_c,R_f)
\label{eq:ase}
\end{equation}
and
\begin{equation}
1 + \frac{ \eta_{cl}} { \eta_0} = f''_2 (r_0,\Xi)
\label{eq:asf}
\end{equation}
\noindent 
With a least square method,  we find the best function 
$\eta = f_2(\nu_c,R_f,r_0,\Xi)$. First, we find an analytical 
relation between the dynamical friction coefficient in the absence of 
clustering $\eta_0$ and the parameters $R_f$ and $\nu$. We obtain:
\begin{equation}
\begin{array}{cll}
 \eta_0  & = & b_1 + b_2 \nu + b_3 R_f + b_4 \nu R_f  + b_5 \nu^2   \\
 &   &
+ b_6 R_f^2
+ b_7  \nu R_f^2 + b_8 \nu^2  R_f^2 + b_9 \nu^2  R_f^2  \\
\label{eq:asg}
\end{array}
\end{equation}
\noindent 
with:
\[
\begin{array} {cllcll} 
 b_1 & = & 1.00655066, \; \;
 b_2 & = & -0.01224778, \\
 b_3   & = &  0.03836632, \; \; 
 b_4 & = & -0.03337465, \\
 b_5 & = & 0.00288929, \; \;
 b_6 & = & -0.02622193, \\
 b_7 & = & 0.04374888, \; \;
 b_8 & = & 0.00690435, \\
 b_9 & = & 0.00860054,  
\end{array}
\]
\noindent 
We perform a $\chi^2$ test between the values of $\eta_0$ obtained
from the last equation and the values of $\eta_0$ calculated integrating
Eq. (\ref{eq:asc}). Our result for the range $1.2  \leq \nu_c \leq 3.2$ 
is excellent: $ \chi^2  \approx 10^{-8}$.\\
We do  the same job for the quantity $f''_2 (r_0,\Xi)$, finding:
\begin{equation}
f''_2 = c_1 + c_2 r_0 + c_3 \Xi + c_4 \Xi r_0
+ c_5  r_0^2 + c_6 \Xi r_0^2 
\label{eq:ash}
\end{equation}
with:
\[
\begin{array}{cllcll}  
c_1&=& 2.26516888, \; \; \; \;
c_2&=&-0.09679056,  \\
c_3&=& 0.00054051, \; \; \; \; 
c_4&=&-0.0001255044, \\
c_5&=& 0.00850948, \; \; \; \;
c_6&=& 0.000009336723 
\end{array} 
\]
\noindent 
An analogue  $\chi^2$ test  gives  $ \chi^2  \approx 10^{-8}$ for the 
range $ 20 \leq \Xi \leq 2 \cdot 10^3$.\\
The function $\eta=f_2 (\nu,R_f,r_0,\Xi)$ is given by the product of 
Eqs. (\ref{eq:asg}) and (\ref{eq:ash}). These contain $54$ terms.
\begin{eqnarray}
\eta & = & f_2 (\nu,R_f,r_0,\Xi)  =  \nonumber \\ 
& = & ( b_1 + b_2 \nu + b_3 R_f + b_4 \nu R_f  + b_5 \nu^2   
+ b_6 R_f^2+ b_7  \nu R_f^2  \nonumber \\
& + & b_8 \nu^2  R_f^2 + b_9 \nu^2  R_f^2) 
\cdot ( c_1 + c_2 r_0 + c_3 \Xi \nonumber \\
& + & c_4 \Xi r_0 + c_5  r_0^2 + c_6 \Xi r_0^2 )  
\label{eq:asmm}  \\ 
\nonumber
\end{eqnarray}
\noindent
However, we have also found an empirical formula with only $13$ terms
that represents a good approximation. We performed a $\chi^2$ test between 
the results obtained from the $ 13$-term equation and the results obtained 
from the numerical integration. The result is $\chi^2 = 2.17 \cdot 10^{-3}$. 
The same test performed on the $ 54$-term equation gives 
$\chi^2 \approx 10^{-8}$. Our $ 13$-term equation reads as:
\begin{equation}
\begin{array}{cll} 
\eta & \approx & d_1 + d_2 \nu + d_3 R_f+ d_4 r_0+d_5 \Xi + d_6 \nu R_f  \\
\; \; \; \; &+& d_7 \nu^2  +  d_8 R_f^2 +d_9 r_0^2 + d_{10} r_0 \Xi  \\ 
\; \; \; \; &+& d_{11} \nu R_f^2 + d_{12} \nu^2 R_f+ d_{13} r_0^2 \Xi   \\ 
\label{eq:asm}
\end{array} 
\end{equation}
\noindent 
with:
\[
\begin{array}{cllcll} 
d_1&=&2.280000, \; \; \; \;
d_2&=&-0.027743, \\
d_3&=&0.086906, \; \; \; \;
d_4&=&-0.097425, \\
d_5&=&5.440520, \; \; \; \;
d_6&=&-0.075599, \\
d_7&=&0.006545, \; \; \; \;
d_8&=&-0.059397, \\
d_9&=&0.008565, \; \; \; \;
d_{10}&=&1.26 \cdot 10^{-4}, \\
d_{11}&=&0.099099, \; \; \; \;
d_{12}&=&0.015640, \\
d_{13}&=&9.39 \cdot 10^{-6}  
\end{array}
\]
\noindent 
The function $\tau=f_1 \circ f_2$ is given by:
\begin{equation}
f_1 \circ f_2  =  a_1 \left[  f_2(\nu,R_f,r_0,\Xi)  \right]^2  + 
a_2 \left[   f_2(\nu,R_f,r_0,\Xi)  \right] +a_3  
\label{eq:asn} 
\end{equation}
\noindent
that is:
\begin{equation}
\tau  =  a_1 \left[ \eta_0 \left( 1 + \frac{\eta_{cl}}{\eta_0} \right) \right]^2  
+ a_2 \left[ \eta_0 \left( 1 + \frac{\eta_{cl}}{\eta_0} \right) \right] +a_3   
\nonumber
\end{equation}
where the values of $a_n$ are given in Sect. 2 and the function $f_2$ is given
by Eq. (\ref{eq:asmm}) or by Eq. (\ref{eq:asm}).\\

\section{A semi-analytical relation between $\tau$ and ${\overline \delta}$}

Our aim is to find a semi-analytical relation 
$\tau = f_3 ({\overline \delta}, \eta)$ for the intervals 
$10^{-4} \le {\overline \delta} \le  10^{-2}$ and 
$0 \le \eta \le 3.1$.\\
The first step is the determination of a function 
$\tau=g({\overline \delta})$ for a fixed value of $\eta$.
We solve Eq. (\ref{eq:p}) for $10^{-4} \le {\overline \delta} \le 10^{-2}$ 
and $\eta = 0.01$ and by a least square method we find the
function $\tau=g({\overline \delta})$:
\begin{equation}
\tau = e_1 {\overline \delta}^2 + e_2 {\overline \delta} + e_3
\end{equation}
\noindent 
with:
\[
\begin{array}{l}
e_1 = 5678.65 , \\
e_2= 45.3429, \\
e_3 = -0.01591, \\
\end{array}
\]
The value of the correlation coefficient between this function and
the numerical integration is $r^2 =1.000$. With this method we 
find $\tau=g({\overline \delta})$ for several values of $\eta$ inside 
the interval  $0 \le \eta \le 3.1$. We can write the dimensionless 
collapse time as the product:
\begin{equation}
\tau=g({\overline \delta}) \cdot h(\eta)
\end{equation}
\noindent
where the function $ h(\eta)$ can be written as
\begin{equation}
h(\eta) = K \cdot \left[ 1 + f(\eta) \right] 
\end{equation}
and, for $10^{-4} \le {\overline \delta} \le 10^{-2}$ and 
$0 \le \eta \le 3.1$, $K$ is constant: $K=-1494.7705898$. 
So:
\begin{equation}
\tau = F ( {\overline \delta}, \eta) = K \cdot \left[ 1 + f(\eta) \right] \cdot
g ({\overline \delta}) 
\end{equation}
A $\chi^2$ test performed between the values obtained for 
$ \tau$ from the numerical integration and the values obtained 
from Eq. (\ref{eq:p}) gives the result: $\chi^2 \sim 10^{-4}$.\\

\section{Conclusions and discussion}

In Sect. 2 of this work we showed in a quantitative way how the collapse 
time $ \tau$ of a shell of baryonic matter made of galaxies and 
substructure depends on some parameters. When one of the parameters
$ \Xi$ or $ R_{f}$ or $ \nu_{c}$ increases, the collapse time grows.
It means that the effects of the presence of dynamical friction 
should be more evident in the outer regions of rich clusters of galaxies. 
Besides, we show how the collapse 
time of an infalling shell increases with decreasing values of $ r_{0}$, 
and becomes very large for $ r_{0} \leq 2 h^{-1}$ Mpc (see Fig. 4). As a 
consequence,  the slowing down of the collapse of an outer shell 
within a cluster of galaxies owing to the dynamical friction is more 
remarkable in the clusters with nucleus of little dimension.\\
~
In Sect. 3 of this paper we give an analytic relation 
that links 
the dimensionless collapse time $ \tau$ with the coefficient of dynamical
friction $ \eta$. This relation is in excellent agreement with the numerical 
integration of Eq. (\ref{eq:p}) for $ 0 \leq  \eta \leq 3.1$ (see Fig. 5).
Then, we find an analytic relation between $ \eta$ and the parameters on 
which it depends for the value $\overline{\delta} = 10^{-2}$ (we remind 
that the effects of the dynamical friction are negligible 
for $\overline{\delta} > 10^{-2}$).  This is Eq. (\ref{eq:asd}), 
that can be considered as a {\it "low order"} approximation to a 
more realistic situation of an outer shell of a cluster of galaxies with
$\overline{\delta} \leq 10^{-2}$. We also find an empirical formula, 
Eq. (\ref{eq:asm}), that is a good approximation  of Eq. (\ref{eq:asmm}). 
Moreover, the dependence of the dimensionless collapse time on 
$ \nu_{c}$,  $r_{0}$, $ \Xi$ and $ R_{f}$  is shown (Eq. (\ref{eq:asn})).\\ 
Finally, we give an analytical relation
that links $\tau$ with $\eta$ and ${\overline \delta}$.
Here, we wish to stress the usefulness of an analytic relation like 
$ \tau = K \cdot [ 1 + (f_1 \circ f_2)] \cdot 
g ({\overline \delta}) $.  
This is a powerful tool to estimate the effect of the 
dynamical friction in the outer regions of clusters of galaxies 
(Gambera et al. in preparation) 
and to compare the observational data with the theoretical ones.
This is a good method to test how important is the role of the dynamical
friction in the collapse of the clusters of galaxies.\\

\begin{flushleft}
{\it Acknowledgements}
\end{flushleft}
 
We are grateful to V. Antonuccio-Delogu for helpful and stimulating
discussions during the period in which this work was performed 
and the the referee Dr. B. S. Ryden for some useful comments.

\end{document}